\documentclass{article}

\usepackage{arxiv}

\usepackage[utf8]{inputenc} %
\usepackage[T1]{fontenc}    %
\usepackage{hyperref}       %
\usepackage{url}            %
\usepackage{booktabs}       %
\usepackage{amsfonts}       %
\usepackage{nicefrac}       %
\usepackage{microtype}      %
\usepackage{lipsum}
\usepackage{graphicx}
\usepackage{caption}
\usepackage{biblatex}
\usepackage{array}

\title{An Improved Relevance Feedback in CBIR}

\author{
  Subhadip Maji\thanks{GitHub Repo: \emph{https://github.com/pidahbus}} \\
  M.Tech QROR-II \\
  Indian Statistical Institute, Kolkata \\
  Kolkata, 700108 \\
  \texttt{qr1705@isical.ac.in} \\
   \And
 Smarajit Bose \\
  Interdisciplinary Statistical Research Unit \\
  Indian Statistical Institute, Kolkata \\
  Kolkata, 700108 \\
  \texttt{smarajit@isical.ac.in} \\
}
\DeclareUnicodeCharacter{2212}{-}
\begin{document}
\maketitle

\begin{abstract}
Relevance Feedback in Content Based Image Retrieval is a method where the feedback of the performance is being used to improve itself. Prior works use feature re-weighting and classification techniques as the Relevance Feedback methods. This paper shows a novel addition to the prior methods to further improve the retrieval accuracy. In addition with all of these, the paper also shows a novel idea to even improve the 0-th iteration retrieval accuracy from the information of Relevance Feedback.

\end{abstract}


\section{Introduction}

Relevance feedback is a method where we take feedback from users after image retrieval and use this feedback to improve the retrieval accuracy after each iteration gradually. There are several relevance feedback methods to account this improvement. In this paper, we use this relevance feedback in a learning way to the system so that as users use this Content Based Image Retrieval System (CBIR) more and more, the average retrieval accuracy increase and the average iteration number to reach the maximum retrieval accuracy decreases. 

There are many Relevance Feedback methods\cite{muller, ortega, Rui, yoshitaka, Zhang2002, zhou} to further improve the precision after first iteration. But, throughout this paper we have used Feature Re-weighting\cite{gitadas} as the elementary Relevance Feedback method to learn the system to further improve the average retrieval accuracy and average relevance feedback iteration number.

\section{What is Relevance Feedback?}
Human perception of image similarity is semantic, task-dependent and subjective. Although content-based methods give promising directions for image retrieval, the retrieval results based on the similarities of pure visual features are not necessarily semantically and perceptually meaningful. In addition, each type of visual feature tends to capture only one aspect of image property and it is usually hard for a user to specify clearly how different aspects are combined. To tackle these problems, relevance feedback, a technique in traditional text-based information retrieval systems, is introduced. 

Relevance feedback is a supervised active learning technique used to improve the effectiveness of information systems. The main idea is to use positive and negative examples given from the user to improve performance of system. For a given query image, the system first retrieves a list of ranked images with respect to a predefined similarity measure. Then, the user selects the retrieved images as relevant (positive examples) or not relevant (negative examples) to the given query image. The system will then refine the retrieval results based on the user feedback and retrieve a new set of images to the user. The key issue in relevance feedback is how to incorporate positive and negative examples given from the user to refine the query and/or to adjust the similarity measure to improve the system retrieval performance. The mechanism is generally applied iteratively.

\section{Feature Reweighting}
In Feature Reweighting\cite{gitadas}, the information obtained from the 0-th iteration (after the user has classified the images as relevant and non-relevant) is used to assign meaningful weights wj to each of the d=100 features (discussed above). The query image is again compared with each image in the database (minus the 20 images already retrieved), but by using a weighted L1 norm distance measure:

\begin{equation}
    D_2 = \sum_{j=1}^{d} w_j|f_{I_j} - f_{Q_j}|
\end{equation}

The images having the least D2 values are returned (1st iteration). The number of images returned in the 1st iteration is Scope - No. of relevant images returned in the 0-th iteration. The user once again classifies these images, the additional information being utilized to modify the weights; and the same process continues for each iteration. For a particular iteration, the no. of images returned is equal to Scope – Total no. of relevant images returned in the preceding iterations. The process continues till six iterations or when the total number of relevant images returned becomes equal to the Scope. 

\subsection{Choice of weights}
An obvious criterion for the choice of weights [8] is that they should be higher for those features which differ significantly between the relevant and non-relevant classes, and thus discriminate well between relevant and non-relevant images; and lower for those features which behave similarly in both the classes.
Let $\sigma_j^{(t)}$ and $\sigma_{rel,j}^{(t)}$ be the standard deviations of $f_j$ over the sets $N_t \cup R_t$ and $R_t$ respectively, where $R_t$ and $N_t$  are respectively the relevant and non-relevant sets at the t-th RF iteration. An intuitive choice of weight for the feature $f_j$ at the (t+1)- th iteration is:

\begin{equation}
    w_j^{(j+1)} = \frac{\sigma_j^{(t)}}{\sigma_{rel,j}^{(t)}}
\end{equation}

If $\sigma_(rel,j)$ becoming zero, the denominator is assigned a small positive value $\Delta$ to avoid computational problems.

Wu and Zhang proposed an efficient way of using both the relevant and non-relevant samples by forming a discriminant ratio which would determine the ability of a feature to separate relevant images from the non-relevant ones. If $F_{rel,j}^{(t)}=\{f_{I,j} I \in R_t \}$ is the collection of the j-th feature of all images in $R_t$, then the dominant range over relevant images at the t-th iteration for the j-th feature component is defined as,

\begin{equation}
    D_j^{(t)} = [min(F_{rel,j}^{(t)}), max(F_{rel,j}^{(t)})]
\end{equation}

The discriminant ratio proposed is:

\begin{equation}
    \delta_j^{(t)} = 1 - \frac{No.of non-relevant images having the j-th  feature in D_j^{(t)}}{|N_t|}
\end{equation}

The value of $\delta_j$ lies between 0 and 1. It is 0 when the j-th feature of all non-relevant images is within the dominant range and thus, no weight should be given for that feature component. On the other hand, when there are no non-relevant images having their jth feature components lying within the dominant range, maximum weight should be given to that feature component ($\delta_j =1$). Based on this, another choice of weights is:

\begin{equation}
    w_j^{(t+1)} = \delta_j^{(t)} * \frac{\sigma_j^{(t)}}{\sigma_{rel,j}^{(t)}}
\end{equation}

\section{Evaluation Metric}
Maji et al.\cite{maji} in their paper used \texttt{Precision} as the evaluation metric.

\begin{equation}
    \texttt{Precision} = \frac{\texttt{Number of relevant images retrieved}}{\texttt{Number of retrieved images}}
\end{equation}

Generally the number of images retrieved by any CBIR method (called the Scope of the method) is a pre-specified positive integer. Precision values are calculated for each image in the database, and these are averaged over all images (in the database). These averages are conventionally plotted for different values of the scope to provide an illustration of the overall retrieval performance of the method. 

However, under relevance feedback, the scenario is slightly different. Here, after the user identifies the relevant and non-relevant images at each iteration, usually a different set (not necessarily disjoint with the earlier set) of images is retrieved in the following iteration due to change in the search criterion. There are several issues involved here. For example, it is not desirable to return the same image (relevant or non-relevant) to the user a second time after retrieval at an earlier iteration. Therefore one should aim to retrieve a new set of images at each iteration, which does not contain any of the images retrieved earlier. Further, if we consider the initial scope (S) as the number of relevant images the user is looking for; it makes sense to retrieve only S-R number of images at every step, where R is the current number of relevant images. Under such considerations, the total number of images to be retrieved changes after every iteration and it is expected to be different for different images. Hence one other evaluation measures are proposed\cite{bose} whose behaviour remains consistent, irrespective of whether RF has been used or not:

\begin{equation}
    \texttt{Retrieval Accuracy} = \frac{\texttt{Number of relevant images retrieved}}{\texttt{Scope}}
\end{equation}

\section{Databases Used}
This database contains 9144 images from 102 categories. The number of images in each category varies from 34 to 800\cite{feifei}.

\section{Improvement Using Relevance Feedback}
This paper is a continuation of the paper by Maji et al.\cite{maji} where they used the pre-trained deep learning features to get the state-of-the-art result. With their method, they achieved an average precision on this dataset to be around 82\%. On the top of their result, we introduced relevance feedback to further improve the retrieval result. 

The features extracted from the CBIR model referred above is high dimensional (n = 1536). So, to tackle the curse of high dimensionality we proceed calculating relevance feedback accuracies taking roughly first 100 principal components. The idea of doing PCA on extracted features is taken from here\cite{maji}.

Once the top 20 images (having the least distance from the query image) have been retrieved by the classical CBIR approach\cite{maji} (0th iteration), these 20 images are presented to the user, and he/she is asked to manually label each of them as “relevant” or “non-relevant” to his/her query.

\begin{figure}[ht]
    \centering
    \includegraphics[scale=0.7]{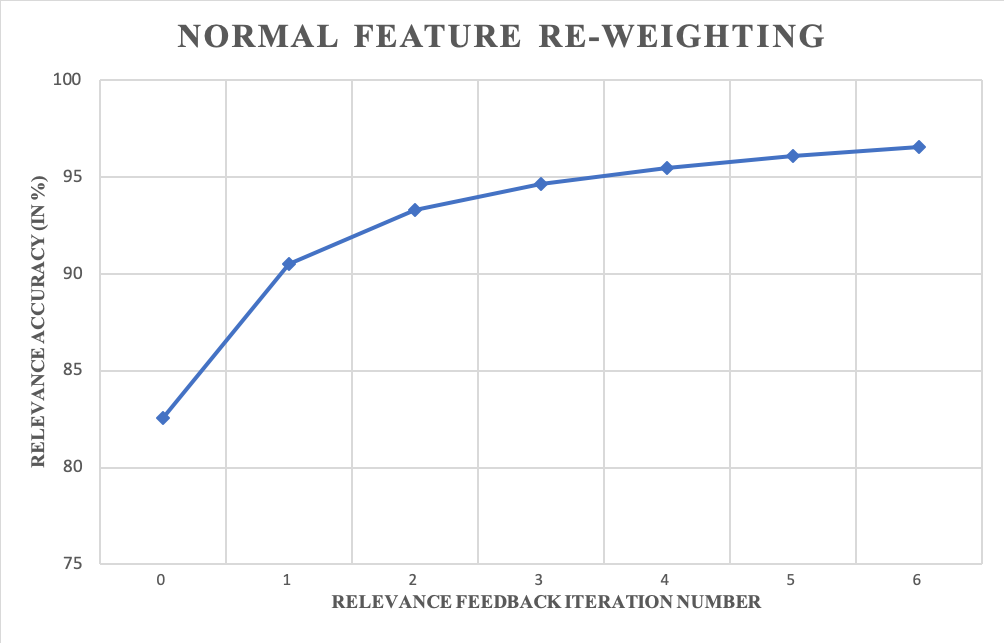}
    \caption{Relevance Accuracies up to iteration 6 for Normal Feature Re-weighting RF Method}
    \label{fig:nfw}
\end{figure}

This feedback is used in subsequent iterations, so that the ranking criterion is updated and a new set of images is retrieved. In this way, the subjective human perception of image similarity is incorporated to the system and the retrieval results are expected to improve, according to the user’s viewpoint, with the RF iterations. The process is generally terminated when there is no further improvement or when the required number of relevant images (here, 20) is retrieved. In this project, we proceed till 6 iterations. (We consider that the user will get tired after 6 iterations and will not like to continue classifying the retrieved images manually even after 6 iterations).The improvement till 6 iterations are shown in the Figure \ref{fig:nfw}.

\section{Further improvement in RF by Grouping Images}
Now we try to further improve in the retrieval accuracy as user continues using the system. This improvement can be done in two ways: Increasing the average retrieval accuracy and decreasing the average iteration number. We did this experiment on the DBCaltech dataset with the total 9144 images. 

\subsection{Experimental Setup}
To implement we divided the DBCaltech dataset in three parts: one image randomly from each group (a total of 102 images) with less precision in RF0 and named it as test images. Retrieval accuracy of these images before and after the experiment will show how much improvement we have got. Another 1000 images randomly selected from the rest 9042 images. These are the images user uses to retrieve similar images over the time. We will show as user uses these images gradually the RF continues to improve. We named these images as validation images. Rests images (a total of 8042) are the dataset from which the similar images will be retrieved. We named it retrieval dataset.

\subsection{Baseline Results}
Without touching validation images, we applied the above described pre-trained neural network method to calculate the precision without RF. The value came to be 56.12\%. Using the Normal Feature Re-weighting method demonstrated above we did the Relevance Feedback retrieval accuracy calculation too. As mentioned above, we fixed the iteration number to be at most 6 for each of the test images. Here along with the retrieval accuracies of the test images we recorded the iteration number at which the accuracy of the image reaches 1.So, 

\texttt{RF iteration number for image i = min\{iteration number at which the accuracy reaches 1, 6\}}

By doing so we get the mean retrieval accuracy to be 82.69\% and average RF iteration number to be 3.71.

\subsection{Procedure}
Now say for a certain period of time user is using the CBIR system we have developed and the images they are using are the validation images which comes outside from the retrieval dataset. Our aim is to classify the images of the retrieval dataset based on the Relevance Feedback provided by the user after each retrieval. 

Now at very first there is no group. After at most 6 iterations for the first validation image, we got $n_1 (n_1 \leq 20)$ relevant images. Now with $n_1$, we create our first group. Now while retrieving images for the second validation image, after first retrieval user selects the relevant images from the sample, if any one of the relevant image matches with the images of the created group we take a random sample equals to the number of irrelevant images and the accuracy becomes 1 after one iteration. If the number of images of the group is less than the number of irrelevant images of that retrieval then we include all the images from that group and rest images we find by Normal Feature Re-weighting. After at most 6 iterations, say we get $n_2$ relevant images. Now we add these $n_2$ images with the previous group. If any of the relevant images after first retrieval of the second validation images does not match with any of the images of the group we try to gather more relevant images by only Norma Feature Re-weighting method and after at most 6 iterations, we create new group with the relevant $n_2$ images. In this process, after at most 6 iterations with the relevant images either we add the images with an existing group or we create a new group.

It may happen that the different relevant images after the first retrieval, came from different groups. In that case we merge all the groups. 

\subsection{Improved Result}
\begin{figure}[ht]
    \centering
    \includegraphics[scale=0.7]{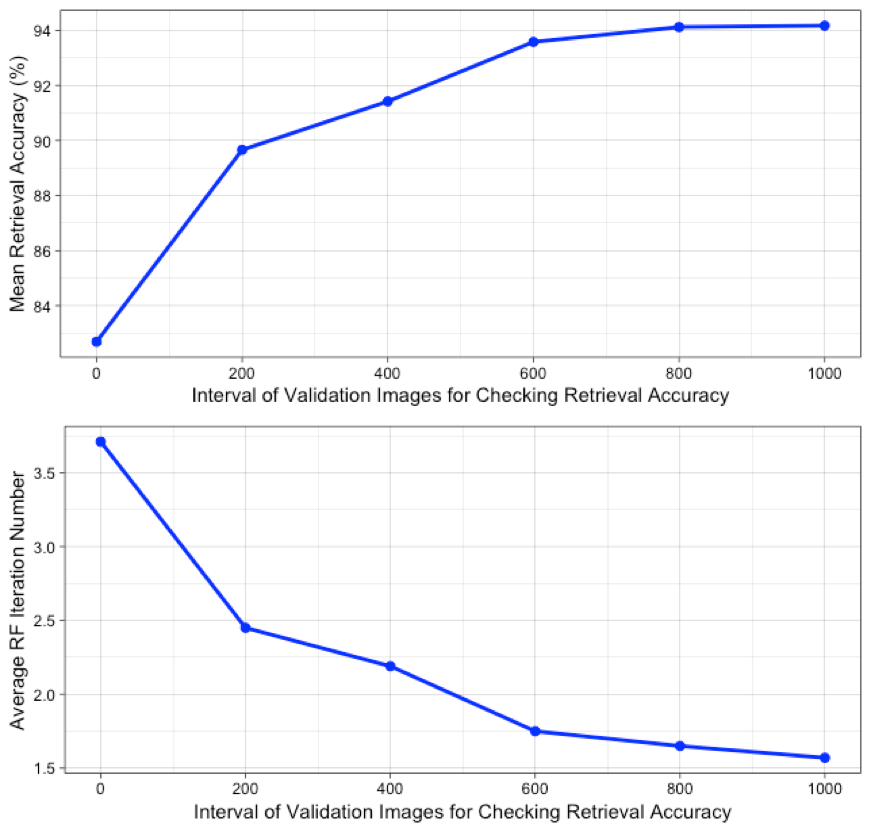}
    \caption{Improvement in Accuracy and RF iteration number over the use of validation images}
    \label{fig:pic2}
\end{figure}

To show the improvement over the use of validation images we selected 5 equal intervals in the process of group formation by the validation images and we checked mean accuracy and RF iteration number at each interval: one after 200 images, second after 400 images and so on. The graphical result is shown in Figure \ref{fig:pic2}.

\section{Improvement of RF0 Precision by Introducing A Classification Model Trained on the Grouped Images}

Here RF0 means the CBIR without any relevance feedback. After doing the above grouping process at the end of 1000 validation images, we get 117 groups, where the original number of groups are 102. It is not hard to believe that if users uses more and more beyond 1000 validation images the number of groups will eventually shrink to 102 from 117. Now the idea is to train a neural network model on these dataset with the specified class and make more appropriate feature set for our dataset only. 

\subsection{Train and Validation dataset}
At first we removed groups with images less than 10. This reduces the group number from 117 to 104. We created two folders: Train and Validation. Randomly pick up 20\% images from each group to the validation folder and rests put up in the train folder. 

\subsection{Procedure}
We have fine-tuned the InceptionResNet model pre-trained on the ImageNet dataset. Firstly, we extracted the convolutional base, then added one two hidden units of 1536 and 256 nodes respectively. Finally added a softmax layer with 104 classes. Compiled the model with RMSprop optimizer with learning rate of 1e-05. We did batch gradient descent with the batch size of 20. At epoch 21 the model gave the highest validation accuracy of around 95\%. Then trained the whole dataset for epoch 21 keeping other things intact. As discussed in the chapter 2, we removed the softmax layer and last layer of 256 dimensions became our new feature representation of the images. Finally we encoded all the images in the retrieved dataset and encoded to 256 dimensional feature vector feeding this newly trained Neural Network model. 

\subsection{Improved Result}
\begin{figure}[ht]
    \centering
    \includegraphics[scale=0.8]{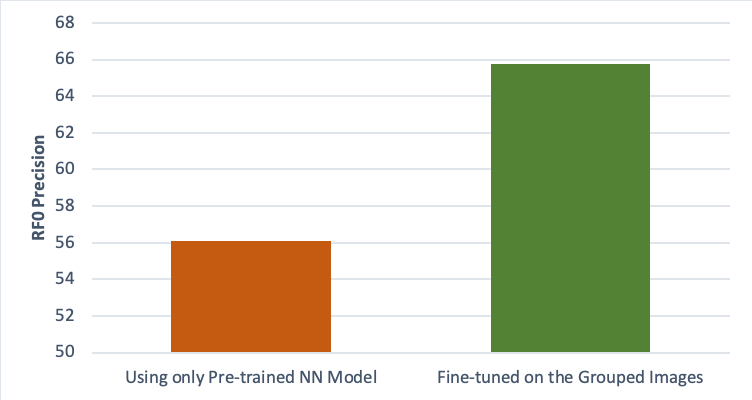}
    \caption{Improvement in RF0 precision over the use of validation images}
    \label{fig:pic3}
\end{figure}

By using the method described by Maji et al.\cite{maji}, for the 102 test images the RF0 precision is 56.12\%. Now after this fine-tuning this result has improved to 65.78\%. The improvement is shown in the Figure \ref{fig:pic3}. The fine-tuned result could be improved further with better network architecture and hyper-parameter tuning. However, the aim of this section is to show the idea of improving the RF0 precision using the relevance feedback from the users.

\section{RF from a Binary Classification perspective}
Like the earlier section this section also concentrates on improving RF0 precision after collecting and storing information during relevance feedback to update model parameters, but unlike grouping images and solving a multi-class classification problem, this section formulates the problem from a simple binary classification approach which ends to be more superior in retrieving relevant images than the earlier methods.

\begin{figure}
    \centering
    \includegraphics[scale=0.6]{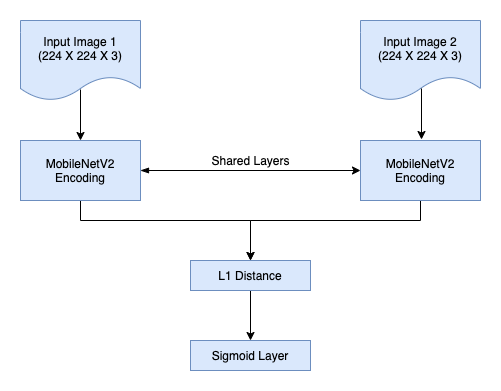}
    \caption{Siamese Neural Network architecture Diagram}
    \label{fig:siamese}
\end{figure}

During relevance feedback users points out given a query image, which images are relevant and which are not. For example, for a scope of $n$, given one query image, user tells that $r_1$ images are relevant and $r_2$ images are not relevant, where $r_1 \geq 0, r_2 \geq 0$ and $r_1 + r_2 = n$. With this information we can create $r_1 \choose 2$ combinations where images in each combination are similar and ${r_1 \choose 1} \times {r_2 \choose 1}$ combinations where images in each combinations are not similar. Each combination consists two images. If we tag similar combinations to be 1 and dissimilar combinations to be 0, then we can formulate this problem as binary classification problem given two images as inputs and finally predicting whether those images are similar or not. We used a Siamese Neural Network\cite{siamese}, to train a model on this dataset. The shared CNN layers in a Siamese Network are taken from the pre-trained layers of MobileNetV2\cite{mobilenetv2} (excluding the last softmax dense layer). These shared CNN layers take two images as input and output two dense vectors (1280 dimension each) for each of the input images. L1 distance is calculated between these two layers. Finally a dense layer of unit 1 with \texttt{sigmoid} activation is used to predict the final output. The model architecture is shown in Figure \ref{fig:siamese}. 

We trained the model with Adam\cite{adam} optimizer using a low learning rate (0.00005), as the model parameters were pre-trained. We spilt the dataset into train and validation set. It took typically 4-5 epochs to train the model. After model training is complete, we extracted the \texttt{MobileNetV2 Encoding} from the model architecture, whose input is an image array (224 X 224 X 3) and output 1280 dimensional vector. We use this tuned encoding layer as our new CBIR encoder which creates feature representation given an image. It turned out that fine-tuning the model encoder with the information we got from relevance feedback improves the quality of the CBIR encoder for retrieval.

\begin{figure}
    \centering
    \includegraphics[scale=0.5]{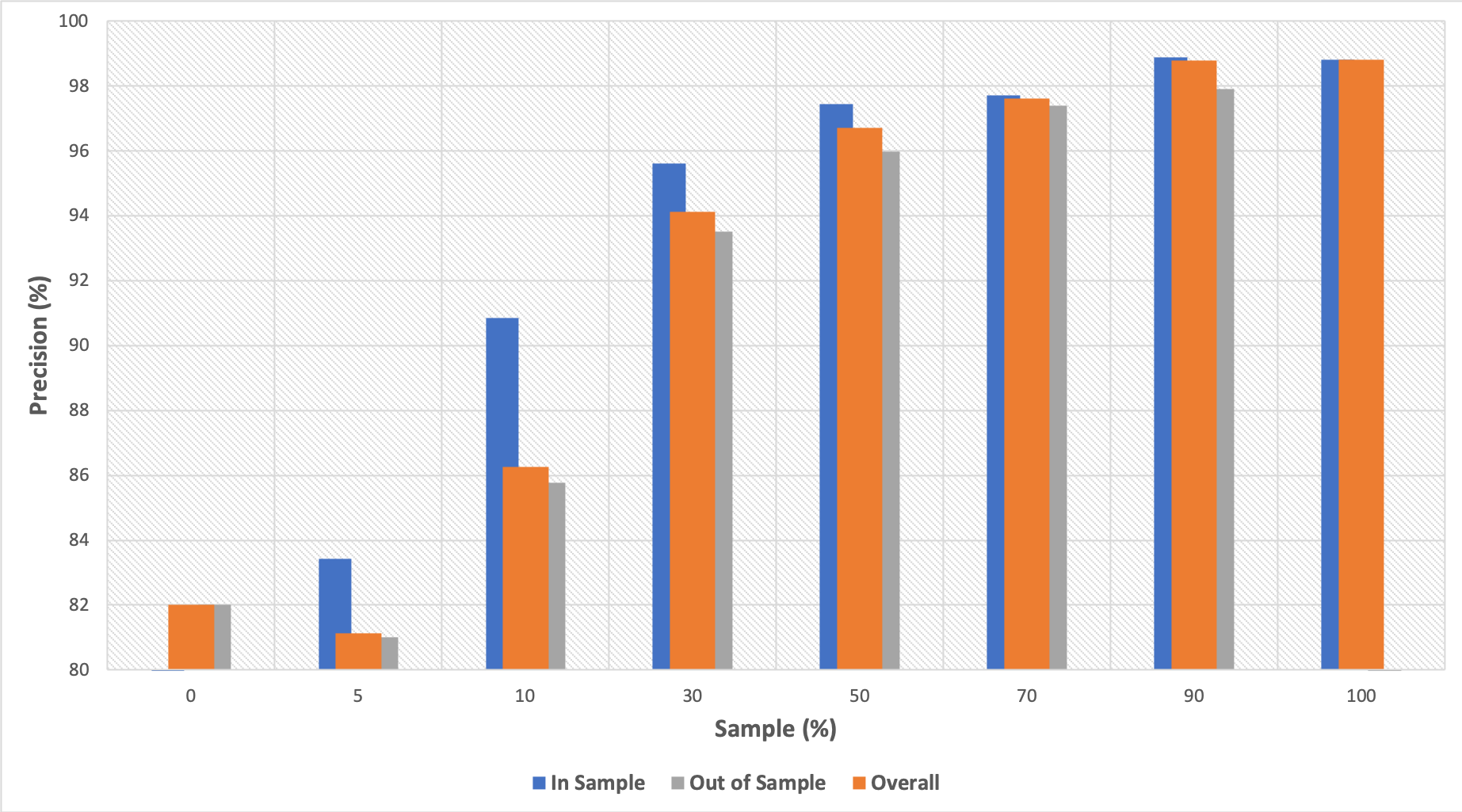}
    \caption{Change in Average Precision values with the change of sample size in Caltech Dataset}
    \label{fig:sample_precision}
\end{figure}

To evaluate the efficacy of the approach we took randomly $x\%$ of the images from the database as query images and retrieved images using the CBIR encoder model described by Maji et al.\cite{maji}. Once the retrieval results are prepared user feedback is used to create the binary class dataset above-mentioned. Here $x \in \{0, 5, 10, 30, 50, 70, 90, 100\}$. We fit separate models and got the \texttt{MobileNetV2 Encoder}for each of the $x$ values. This newly tuned encoder models are used to calculate precision onward. Table \ref{tab:sample_precision} and Figure \ref{fig:sample_precision} show the change of Average precision values with the change of sample ($x$) size from 0 to 100 using new encoder models. In sample precision means the average precision while the $x\%$ sampled images are used as query images and Out of Sample Precision means the average precision while the $(100-x)\%$ images are used as query images. Overall Precision is the average precision of all the images in Caltech dataset. Sample ($x$) of 0\%, means that in this case no image is used in the relevance feedback and 82.02\% precision is the precision value using the method described by Maji et al.\cite{maji}. It is clearly seen that as we increase the sample size, except for 5\%, all precision values increased significantly using the new encoder models we got from the binary classification approach. 

Finally from the practical point of deploying a CBIR system, we suggest a better road map to follow. First deploy the system using the processes informed by Maji et al.\cite{maji}. Store the relevance feedback information of the user. Once it is found that around 10-15\% time the CBIR system has been used, use the stored feedback information to build a Siamese Network model. Once the model is build extract the encoder model and replace it with the one from Maji et al.\cite{maji}. As shown above, it will improve the retrieval precision. One is suggested to update this model encoder after certain time interval as more user feedback information is collected. 

Two points to be noted here:
\begin{itemize}
    \item Above-mentioned 10-15\% means, if there are 1000 images in the database, then after 100-150 iterations of retrieval it is recommended to build the new encoder model.
    \item In the whole paper we used the database images as query images, but from the practical usability perspective, the query image may come outside the CBIR database also. 
\end{itemize}

\begin{table}
    \centering
    \begin{tabular}{|c|>{\centering\arraybackslash}p{3cm}|>{\centering\arraybackslash}p{3cm}|>{\centering\arraybackslash}p{3cm}|}
        \hline
        \textbf{Sample (\%)} & \textbf{Overall Precision (\%)} &  \textbf{In Sample Precision (\%)} & \textbf{Out of Sample Precision (\%)} \\
        \hline
        0 & 82.02 & - & 82.02 \\
        \hline
        5 & 81.12 & 83.42 & 81.00 \\
        \hline
        10 & 86.27 & 90.84 & 85.76 \\
        \hline
        30 & 94.13 & 95.61 & 93.51 \\
        \hline
        50 & 96.71 & 97.45 & 95.97 \\
        \hline
        70 & 97.61 & 97.71 & 97.39 \\
        \hline
        90 & 98.78 & 98.88 & 97.90 \\
        \hline
        100 & 98.82 & 98.82 & - \\
        \hline
    \end{tabular}
    \caption{Change in Average Precision values with the change of sample size in Caltech Dataset}
    \label{tab:sample_precision}
\end{table}

\section{Conclusion}
This paper shows a novel method to gradually improve the retrieval accuracy of CBIR collecting the user feedback by grouping the images and also uses this feedback information to further improve the 0-th iteration retrieval accuracy. 

\clearpage
\printbibliography
\end{document}